\documentclass[a4paper,11pt]{article}
\usepackage{pos}
\usepackage{gensymb}
\usepackage{caption}
\usepackage{float}

\title{Methods for the suppression of background cascades produced along atmospheric muon tracks in the Baikal-GVD}
\ShortTitle{Suppression of background cascades in the Baikal-GVD}

\author[a]{V.A.~Allakhverdyan}
\author[b]{A.D.~Avrorin}
\author[b]{A.V.~Avrorin}
\author[b]{V.M.~Aynutdinov}
\author[c]{R.~Bannasch}
\author*[d]{Z.~Barda\v{c}ov\'{a}}
\author[a]{I.A.~Belolaptikov}
\author[a]{I.V.~Borina}
\author[a,1]{V.B.~Brudanin}
\author[e]{N.M.~Budnev}
\author[a]{V.Y.~Dik}
\author[b]{G.V.~Domogatsky}
\author[b]{A.A.~Doroshenko}
\author[a,d]{R.~Dvornick\'{y}}
\author[e]{A.N.~Dyachok}
\author[b]{Zh.-A.M.~Dzhilkibaev}
\author[d]{E.~Eckerov\'{a}}
\author[a]{T.V.~Elzhov}
\author[f]{L.~Fajt}
\author[g,1]{S.V.~Fialkovski}
\author[e]{A.R.~Gafarov}
\author[b]{K.V.~Golubkov}
\author[a]{N.S.~Gorshkov}
\author[e]{T.I.~Gress}
\author[a]{M.S.~Katulin}
\author[c]{K.G.~Kebkal}
\author[c]{O.G.~Kebkal}
\author[a]{E.V.~Khramov}
\author[a]{M.M.~Kolbin}
\author[a]{K.V.~Konischev}
\author[h]{K.A.~Kopa\'{n}ski}
\author[a]{A.V.~Korobchenko}
\author[b]{A.P.~Koshechkin}
\author[i]{V.A.~Kozhin}
\author[a]{M.V.~Kruglov}
\author[b]{M.K.~Kryukov}
\author[g]{V.F.~Kulepov}
\author[h]{Pa.~Malecki}
\author[a]{Y.M.~Malyshkin}
\author[b]{M.B.~Milenin}
\author[e]{R.R.~Mirgazov}
\author[a]{D.V.~Naumov}
\author[a]{V.~Nazari}
\author[h]{W.~Noga}
\author[b]{D.P.~Petukhov}
\author[a]{E.N.~Pliskovsky}
\author[j]{M.I.~Rozanov}
\author[a]{V.D.~Rushay}
\author[e]{E.V.~Ryabov}
\author[b]{G.B.~Safronov}
\author[a]{B.A.~Shaybonov}
\author[b]{M.D.~Shelepov}
\author[a,d,f]{F.~\v{S}imkovic}
\author[a]{A.E. Sirenko}
\author[i]{A.V.~Skurikhin}
\author[a]{A.G.~Solovjev}
\author[a]{M.N.~Sorokovikov}
\author[f]{I.~\v{S}tekl}
\author[b]{A.P.~Stromakov}
\author[a]{E.O.~Sushenok}
\author[b]{O.V.~Suvorova}
\author[e]{V.A.~Tabolenko}
\author[e]{B.A.~Tarashansky}
\author[a]{Y.V.~Yablokova}
\author[c]{S.A.~Yakovlev}
\author[b]{D.N.~Zaborov}

\affiliation[a]{Joint Institute for Nuclear Research, Dubna, Russia}
\affiliation[b]{Institute for Nuclear Research, Russian Academy of Sciences, Moscow, Russia}
\affiliation[c]{EvoLogics GmbH, Berlin, Germany}
\affiliation[d]{Comenius University, Bratislava, Slovakia}
\affiliation[e]{Irkutsk State University, Irkutsk, Russia}
\affiliation[f]{Czech Technical University in Prague, Prague, Czech Republic}
\affiliation[g]{Nizhny Novgorod State Technical University, Nizhny Novgorod, Russia}
\affiliation[h]{Institute of Nuclear Physics of Polish Academy of Sciences (IFJ~PAN), Krak\'{o}w, Poland}
\affiliation[i]{Skobeltsyn Institute of Nuclear Physics, Moscow State University, Moscow, Russia}
\affiliation[j]{St.~Petersburg State Marine Technical University, St.Petersburg, Russia}

\note{Deceased.}

\emailAdd{bardacova@fmph.uniba.sk}

\abstract{The Baikal-GVD (Gigaton Volume Detector) is a $\rm{km}^3$- scale neutrino telescope located in Lake Baikal.  Currently (year 2021) the Baikal-GVD is composed of 2304 optical modules divided to 8 independent detection units, called clusters. Specific neutrino interactions can cause Cherenkov light topology, referred to as a 	cascade. However,  cascade-like events originate from discrete stochastic energy losses along muon tracks. These cascades produce the most abundant background in searching for high-energy neutrino cascade events. Several methods have been developed, optimized, and tested to suppress background cascades.}

\FullConference{37$^{\rm{th}}$ International Cosmic Ray Conference (ICRC 2021)\\
	July 12th -- 23rd, 2021\\
	Online -- Berlin, Germany}

\begin{document}
	\maketitle

	\section{Introduction}
	
	One of the main goals of the Baikal-GVD (Gigaton Volume Detector)  is to investigate sources of astrophysical neutrinos in energy range from 1 TeV up to 100 PeV. The Baikal-GVD is currently under construction, located in the southern part of Lake Baikal 3-4 km from the shore \cite{GVD}.  
	
\begin{figure}[h]
		\centering
		\includegraphics[scale=0.29]{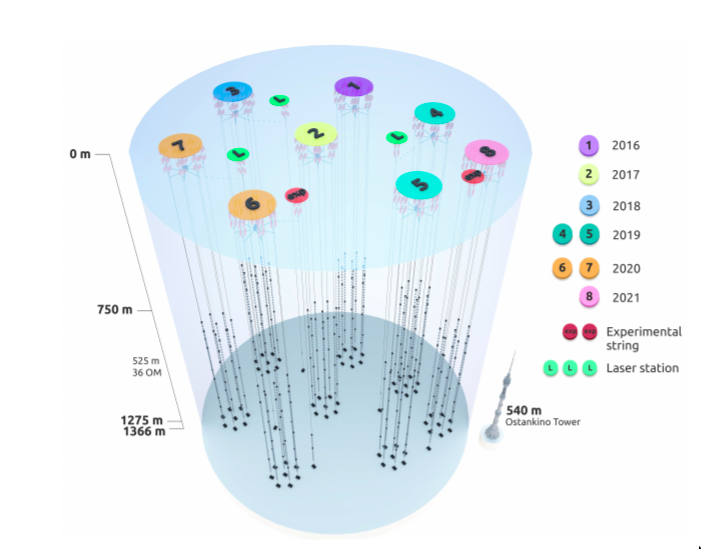}
		\includegraphics[scale=0.18]{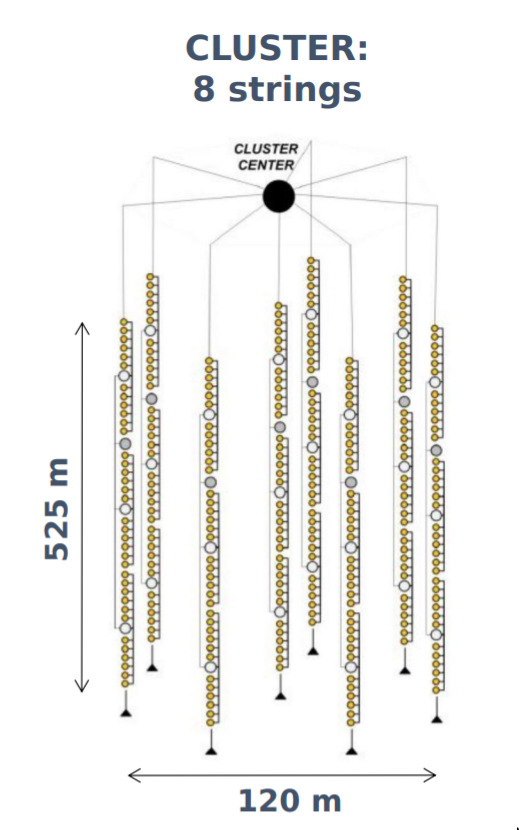}
		\caption{Left: Illustration of the Baikal-GVD detector in year 2021. It contains 8 clusters, 3 laser stations and 2 experimental strings. Right: Sketch of the Baikal-GVD cluster.}
		\label{GVDsketch}
	\end{figure}

	The detector is a three dimensional array of Optical Modules (OMs), which are able to detect Cherenkov light of secondary charged particles produced in neutrino interactions. The OMs are attached to the vertical strings  at depths from 750 m to 1275 m. The whole detector is sub-arranged into functionally independent units, referred to as clusters (see Fig. \ref{GVDsketch}). Each cluster comprises 288 OMs attached to  8 strings.  The effective volume of one cluster for neutrino cascade events with energy above 100 TeV is $\approx 0.05 ~\rm{km}^3$
	\cite{clusterVolume}.
	At the current stage (year 2021), the Baikal-GVD is composed of  8 clusters, comprising 2304 OMs with a total effective volume $\approx$ 0.4 $\rm{km}^3$. 
	
	According to the Cherenkov light topology of the secondary charged particles, there are different types of neutrino events. Muons created by muon neutrino charged current interactions passing through the water leave behind a track. These muons can be both astrophysical or atmospheric origin. Moreover, atmospheric muons may also reach the Earth's surface in a bundle. Charged-current interactions of electron neutrino and low-energy tau neutrino, neutral-current interactions of all three neutrino flavours create a  light
	signature of a single cascade. The most energetic cosmic neutrino events were detected as cascades (in IceCube experiment \cite{IceCubeCascade}).  However, the most abundant background present in the cascade channel originate from discrete stochastic energy losses produced along muon tracks (bremsstrahlung, photonuclear processes, direct electron-positron pair production). 
	
	In this paper, the optimised methods for the suppression of the background cascades that are produced along muon tracks are presented. Majority of these suppression methods are based on search of  hits from the muon track in an event. Optimization of the suppression methods is performed with Monte Carlo (MC) simulations for season 2019. Subsequently, experimental data from season 2019, which corresponds to dataset used in this work in period  between April 2019 and March 2020 were processed with cascade reconstruction software  in parallel \cite{parallel} and  then analyzed. The presented analysis is performed only with single-cluster data.
	\section{Suppression Methods - Monte Carlo Samples}
	\label{Sup}
	To distinguish between cascades induced in neutrino interactions and background cascades produced along muon tracks, different kinds of suppression methods have been developed, tested, and optimized. The performance of the methods to separate signal from background cascades using various MC datasets is described. Especially, their effect has been studied for atmospheric muon bundles sample ($\mu_{\rm{atm}}$), upgoing atmospheric  muon neutrinos ($\nu^{\rm{atm}}_{\mu}$) and atmospheric electron neutrinos ($\nu^{\rm{atm}}_e$). Firstly, only well-reconstructed  contained cascades  have been used for the following analysis. A lower likelihood value of the event means  to be a well-reconstructed cascade. A contained event has its  position  reconstructed within the instrumented detector volume.
	
\begin{figure}[h]
		\centering
		\includegraphics[scale=0.28]{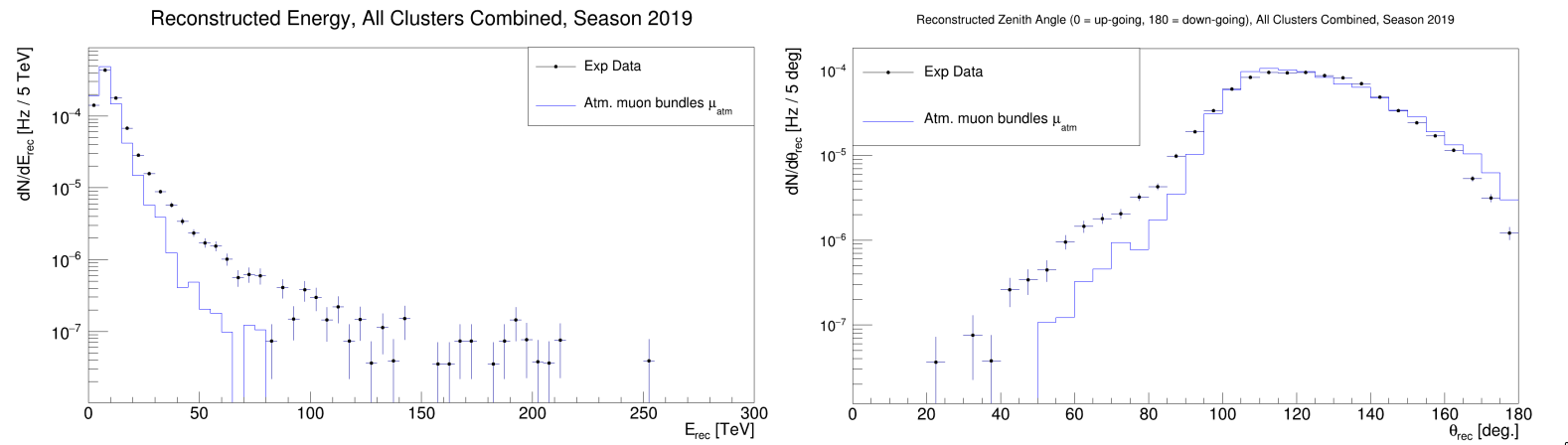}
		\caption{Left: Reconstructed energy  distribution of well-reconstructed contained cascades.  Black points with statistical error bars correspond to experimental data in season 2019 with 5 clusters combined together. The MC prediction for background cascades (scaled by a factor 1.193) is shown in blue. Right: Zenith angle distribution, an angle in the interval ($0\degree,90\degree$) is for upgoing events, while the interval ($90\degree,180\degree$) corresponds to downgoing events.}
		\label{energy}
	\end{figure}

	In particular, the focus is put mainly on the suppression of background cascades from atmospheric muons, which constitute a major part of reconstructed cascades from experimental data, see Tab.\ref{TableExpData}. The effective livetime of the experimental dataset for season 2019 is $\approx$ 1557 days (for all 5 clusters combined). The last row of the Tab.\ref{TableExpData} shows the ratio of the number of experimental well-reconstructed contained events to the number of events expected from MC  $\mu_{\rm{atm}}$ simulation (using the same exposition time and reconstruction analysis).  Their reconstructed zenith angle and energy distributions  for season 2019 are shown in Fig. \ref{energy}. Most of the downgoing events from the experimental data is compatible with the MC $\mu_{\rm{atm}}$ expectation.  Below, the description of the suppression methods is presented.
	
	\begin{center}
		\captionof{table}{Experimental data in the whole season 2019 and MC prediction for $\mu_{\rm{atm}}$ background cascades.} 
		\label{TableExpData}
		\begin{tabular}{||c c c c c c ||} 
			\hline
			Season 2019&Cluster1 & Cluster2 & Cluster3 & Cluster4 &Cluster5\\ [0.5ex] 
			\hline\hline
			Exposition time [days] & 	299.131&	330.043&	320.948&	299.623&	307.322	\\ 		
			No. Reconstructed Cascades [\#]& 10574 &11525&16420
			&17062&16211\\
			No. Contained+Well-Reco Cascades [\#]&3548 & 3843& 5666& 5510 &5367\\
			MC $\mu_{\rm{atm}}$ Expectation [\#]&3403&3457&4641&4520&4004\\
			Normalization Factor& 1.043&1.111 &1.221& 1.219 & 1.341\\
			\hline		
		\end{tabular}
	\end{center}
	
	\subsection*{nTrackHits Filter} 
\begin{figure}[h]
		\centering
		\includegraphics[scale=0.3]{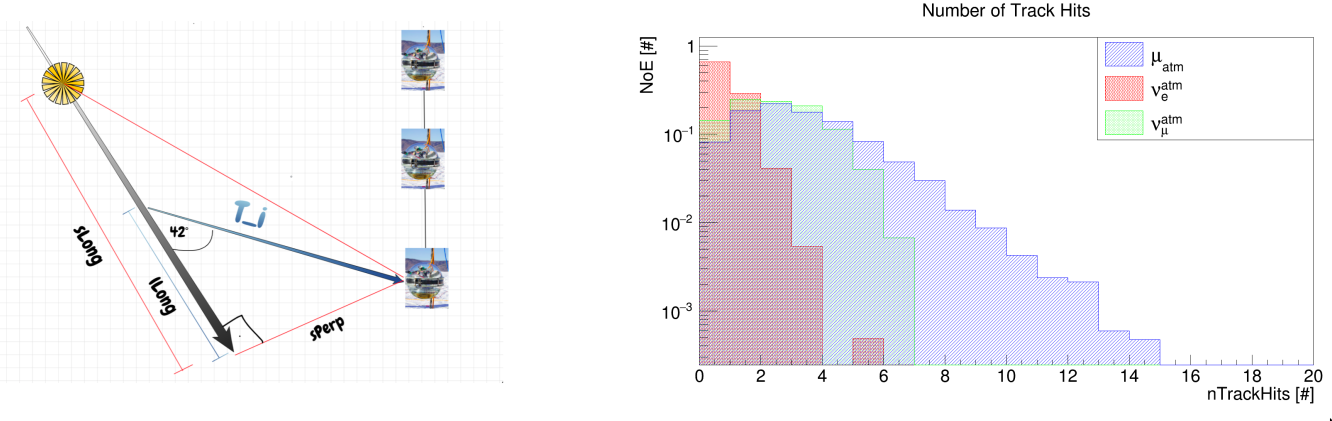}
		\caption{Left: Illustration of the background cascade geometry used for the $T_i$ calculation. Right: Distribution of found nTrackHits for MC datasets.}
		\label{trackSketch}
	\end{figure}

	The goal of the nTrackHits method is to count the number of hits originating from the muon (referred to as nTrackHits) and also to sum their charges (trackCharge). The track hit detection time $t_i$ on a given OM has to fulfill the  condition:
	\begin{equation}
		|t_i - T_i| < \delta t,
	\end{equation}
	where $\delta t = 20 ~\rm{ns}$ is an extra time window and $T_i$ is expected hit time of detection on the OM from the muon track, calculated  according to equation: 
	\begin{equation}
		T_{i} = t_{\rm{recoCascade}} + \rm{(sLong - lLong)}\cdot \frac{1}{c} + \sqrt{\rm{sPerp}^2 + \rm{lLong}^2}\cdot \frac{1}{c_w},
	\end{equation}
	where $t_{\rm{recoCascade}}$ is the reconstructed cascade time, $c$ is the speed of light in vaccum (i.e. approximately equal to muon velocity), $c_w$ is  the speed of light in water. See Fig. \ref{trackSketch} (left) for the meaning of the other variables in Eq. 2. The nTrackHits is calculated in each muon direction given  by iterative seaching over zenith  and azimuth angle in the whole $4 \pi$ geometry. Position and time of the reconstructed cascade fix the track in a single point. The direction in which the most nTrackHits were found is selected for further analysis. Reconstructed cascade direction cannot be used for the track, because it can be biased by  track hits incidentally used in the reconstruction. Fig. \ref{trackSketch} (right) shows the distribution of nTrackHits for above-mentioned datasets. As we would expect, for the  $\mu_{\rm{atm}}$  the highest numbers of nTrackHits were identified. For signal cascades from $\nu^{\rm{atm}}_e$ sample, with no muon track present in it, a few non-zero nTrackHits were falsely identified due to noise hits contamination. However, a cut on nTrackHits can provide separation between signal and background cascades.
	
	\subsection*{BranchRatio Filter}
\begin{figure}
		\centering
		\includegraphics[scale=0.3]{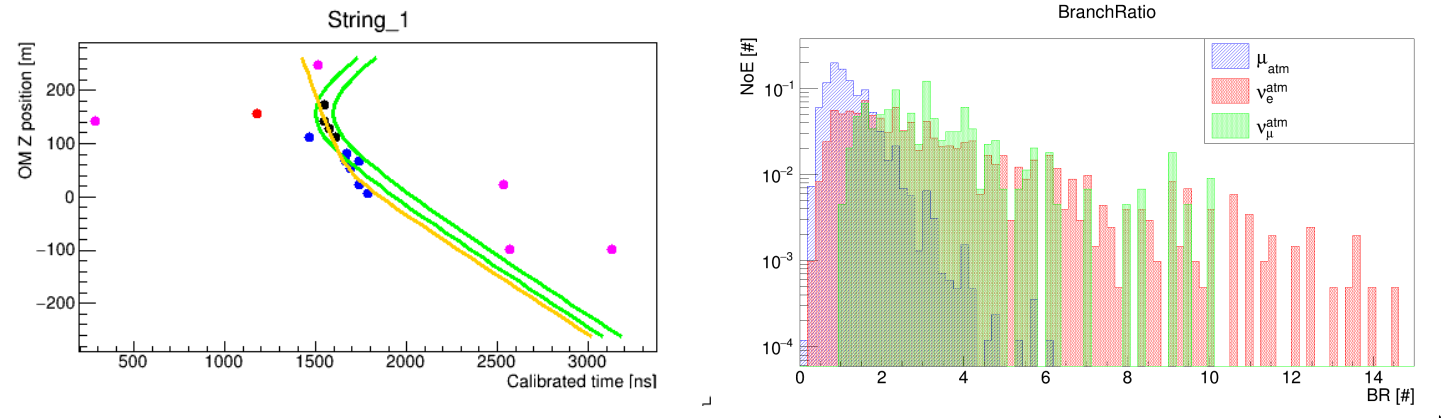}
		\caption{Left: The scatter plot of hit detection time and $z$ coordinate of hit OMs in one of the strings.  Reconstructed cascade is shown by red color, noise hits are given by pink, track hits by blue and  black points correspond to cascade hits. Green band determines the expected time region for the cascade hits and orange line  is expected detection time for track hits. Right: The distributions of $BR$ for three  MC datasets.}
		\label{BR}
	\end{figure}

	The next applied selection criterion, called BranchRatio filter, is used to reduce primarily  downgoing $\mu_{\rm{atm}}$ events. The BranchRatio $BR$ is defined as the ratio of  number of hit OMs that lie in the upper half-plane to the ones in the lower half-plane: $BR = \frac{\rm{nHitsOMs_{\rm{upper}}}}{\rm{nHitsOMs}_{\rm{lower}}}$. The plane
	intersection is determined by the $z$ coordinate of the reconstructed cascade and is perpendicular to vertical cluster strings. For better visualization of the $BR$ calculation, the dependence of hit detection time on the $z$ coordinate of the OM, which detected the hit for the MC downgoing $\mu_{\rm{atm}}$ event is given  in Fig. \ref{BR} (left). Various colors of points are associated with different hits origins. Fig. \ref{BR} (left) shows that most of the hits are located in the lower half-plane, therefore, $BR$ for downgoing $\mu_{\rm{atm}}$ event is assumed to take values lower than 1.  The $BR$ distributions  are displayed in Fig. \ref{BR} (right).
	For signal cascades, the higher values of BR are obtained. The reason is that upgoing events are also present in the $\nu^{\rm{atm}}_e$ sample in contrast to  $\mu_{\rm{atm}}$ events.
	
	\subsection*{CloseHits and QEarly Filter}
\begin{figure}
		\centering
		\includegraphics[scale=0.3]{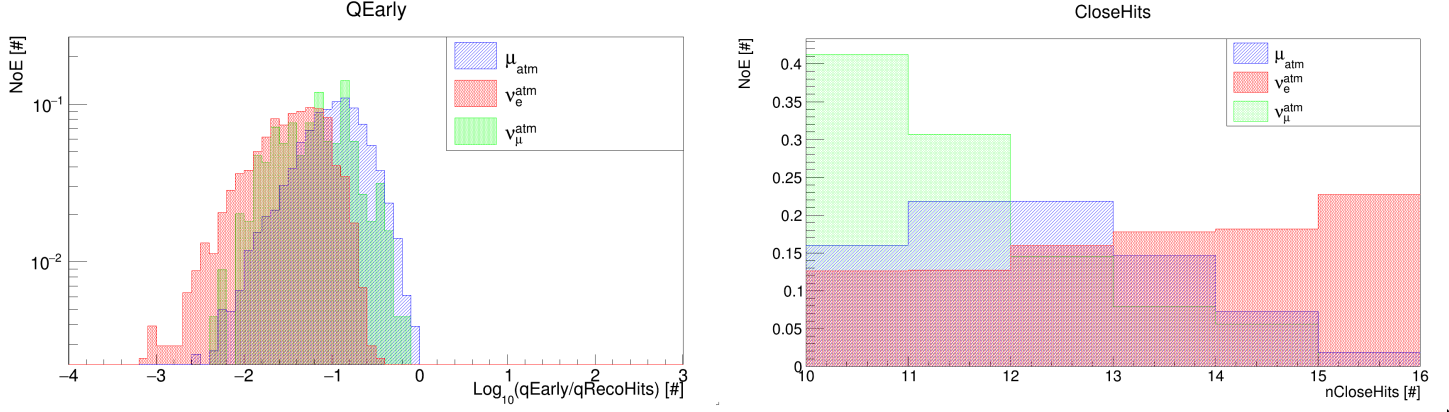}
		\caption{Left: The distribution of the QEarly variable.  QEarly can provide separation between signal and background cascades to some extent. Right: CloseHits distribution for the MC datasets.}
		\label{closeHits}
	\end{figure}

	The CloseHits filter controls how many of the 15 closest OMs w.r.t to the final reconstructed cascade position are hit. For cascade-like events, which follow almost spherical light pattern, $\approx$ 15 nCloseHits are expected to achieve, see Fig. \ref{closeHits} (right). However, as the intensity of light decreases with the distance
	from the reconstructed cascade position and also some of the OMs
	do not have to be in operation, then the cascade does not enlighten all the 15 pinpointed OMs. 
	
\begin{figure}
		\centering
		\includegraphics[scale=0.14]{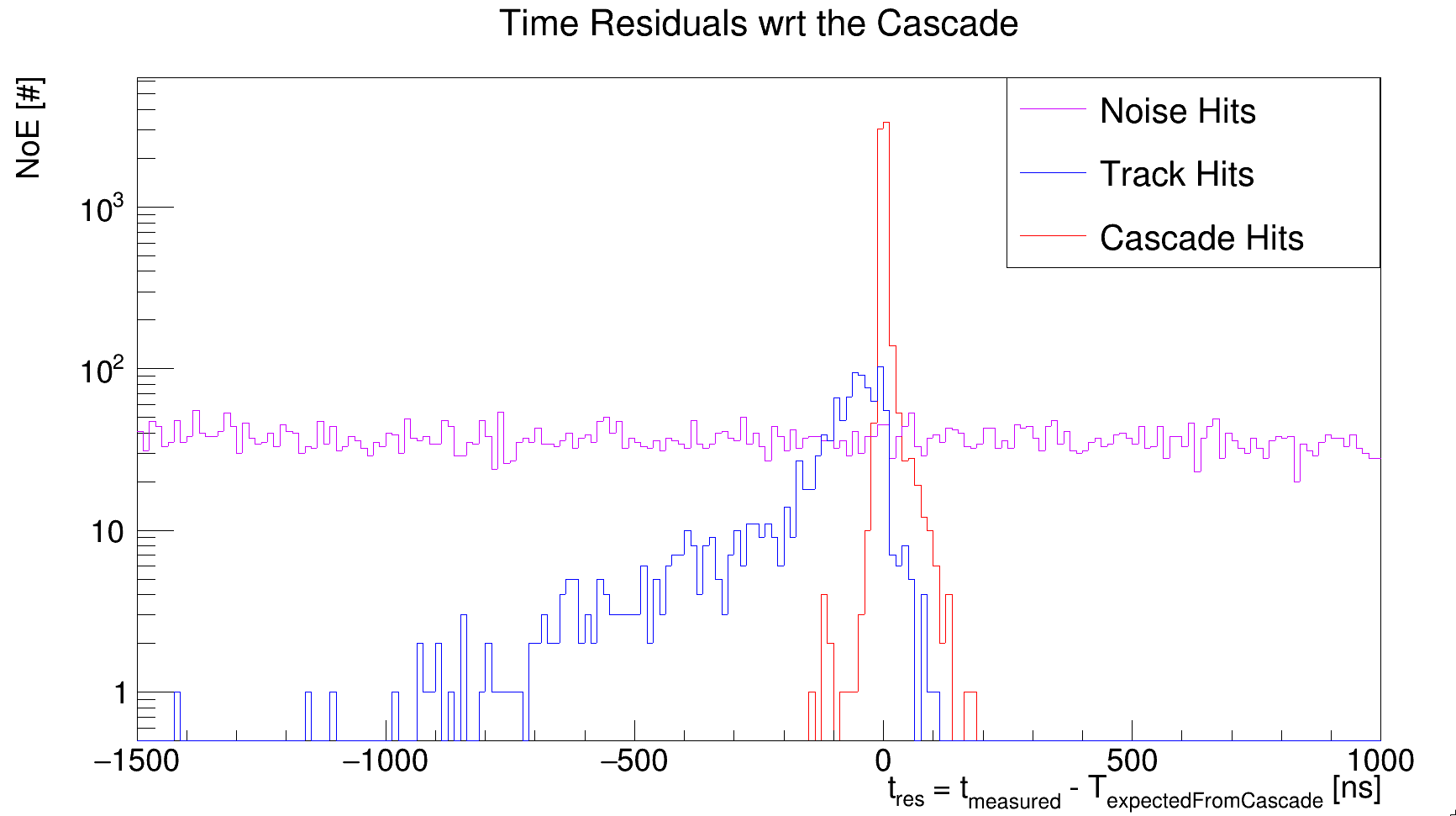}
		\caption{Time residuals $t_{res}$ for MC $\mu_{\rm{atm}}$ event. The hits due to track, cascade, and  randomly distributed noise are shown with blue, red, and pink lines, respectively.}
		\label{timeRes}
	\end{figure}

	The final filter, called QEarly, is inspired by ANTARES \cite{ANTARES}. When background cascade is produced along the muon track, hits coming from the muon are mostly detected earlier at a given OM than it is expected for cascade hits. Fig. \ref{BR} shows  some track hits (blue points) are located before the expected time region for cascade hits (green band). For that reason, QEarly variable is calculated according to the following expression: $\rm{QEarly} = log_{10}(\frac{\rm{qEarly ~[p.e.]}}{\rm{qRecoHits ~[p.e.]}})$, where qEarly is  the charge summed over all hits with a time residual $t_{res}$ w.r.t. the reconstructed cascade: $ -1000 < t_{res}/ns < -40$ and qRecoHits corresponds to the charge summed over expected time for cascade hits (green band): i.e. $ -20 < t_{res}/ns < +20$, see Fig. \ref{timeRes}. As shown in Fig. \ref{closeHits} (left),  muon-like events are  slightly shifted to higher values of the charge qEarly produced by track hits present in $\mu_{\rm{atm}}$ events.

	\section{Boosted Decision Tree and Results}
\begin{figure}
		\centering
		\includegraphics[scale=0.3]{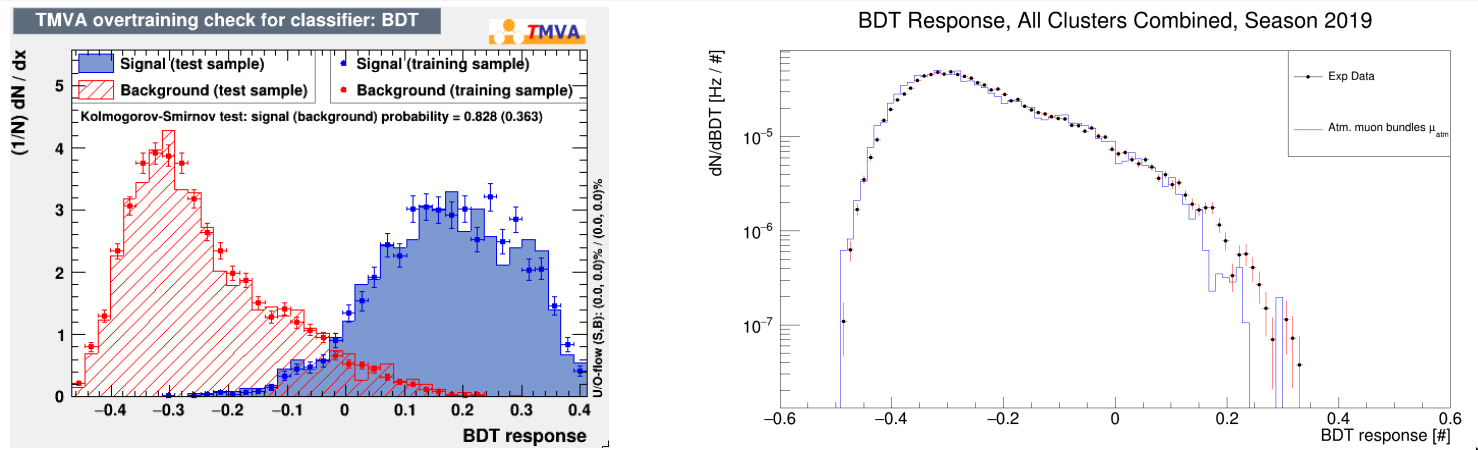}
		\caption{Left: The distribution of BDT response for MC signal (blue) and background cascades (red). Right: The BDT value shown for MC $\mu_{\rm{atm}}$ backgound cascades (blue) and experimental data in season 2019 (black points).}
		\label{BDT}
	\end{figure}

	Any of the developed variables does not give a clear distinction between signal and background cascades. Therefore, to take the most of the discriminating power of many variables,  Boosted Decision Trees (BDTs) (from TMVA ROOT package) were used in the analysis. The BDTs were trained using the 7 variables   as input: nTrackHits, trackCharge, nCloseHits, BranchRatio, QEarly, Z reconstructed cascade coordinate and $\chi^2_{\rm{recoPos}}$ after final position reconstruction. For BDT training and testing  of contained well-reconstructed cascades, we used $\sim$ 3000 and 1000 MC $\nu^{\rm{atm}}_e$ signal cascades, respectively. In case of $\mu_{\rm{atm}}$ sample, almost  6500 and 1900 background cascades were used. After  training and testing, a ranking of input variables is computed. The most important input to the BDTs is a $z$ coordinate of the event. The second most important variable for training is QEarly and then BranchRatio.  After the BDTs are trained and tested , a BDT response for the classification of signal and  $\mu_{\rm{atm}}$ background event  is produced. Fig. \ref{BDT} (left) shows the distribution of the BDT output value for MC signal and  $\mu_{\rm{atm}}$ background cascades. Good agreement of the BDT response for MC $\mu_{\rm{atm}}$ events and experimental data from season 2019 is seen in Fig. \ref{BDT} (right). We also applied the same procedure for uncontained events (horizontal distance of the reconstructed vertex is in interval $(60,100)~\rm{m}$ from the cluster center).

\begin{figure}
		\centering
		\includegraphics[scale=0.29]{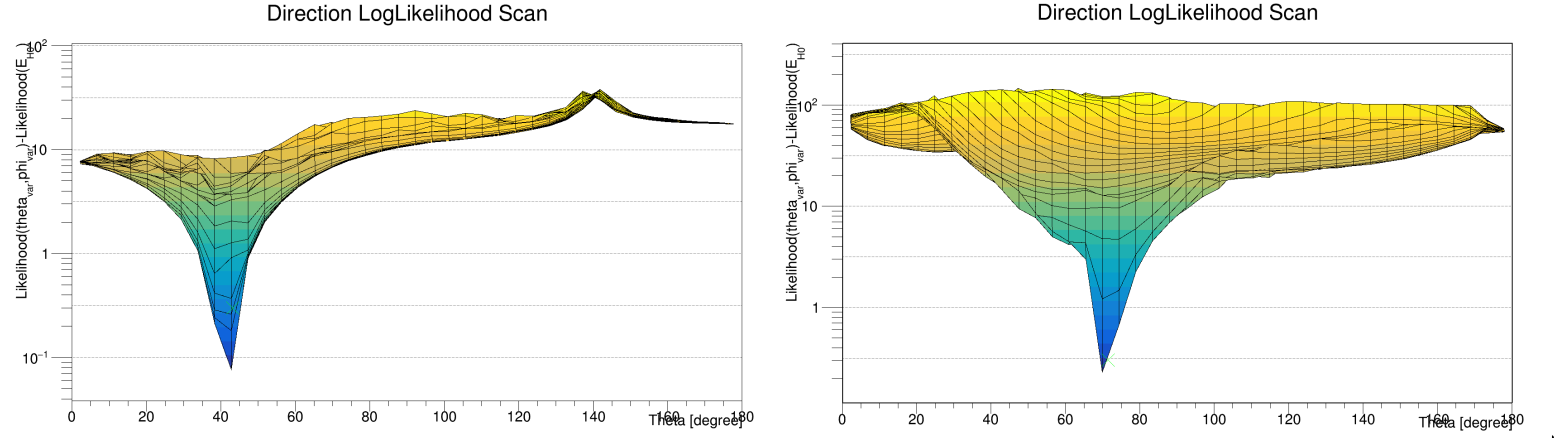}
		\caption{Left: Dependence of the likelihood scan values on the zenith angle for uncontained event with energy $46.6$ TeV. The one significant minimum of the likeihood values is reached for the reconstructed zenith angle. Right: Likelihood scan for contained event with energy $ 79.7$ TeV.}
		\label{scan}
	\end{figure}

	Subsequently, to suppress background cascades, a cut on the BDT value is applied. For signal to background ratio: 1:1000, the maximum significance is obtained when cutting the BDT value at  $\sim$ 0.29. A different BDT cut was determined for uncontained events. Once the BDT cut is applied, 25  contained and uncontained events  remained.  However, most of the  reconstructed events have low energies ($< 20 ~\rm{TeV}$) except two events. Both of them are reconstructed as upgoing with energies $79.7$ TeV for contained event and $46.6$ TeV for uncontained, see Tab. \ref{events}. The likelihood-direction scan for these two events is displayed in Fig. \ref{scan}.

	\begin{center}
		\captionof{table}{Reconstructed parameters of two most energetic events: Cluster, Energy, Zenith angle, Azimuth angle, Horizontal distance, Likelihood, Total charge, Total number of hits, Number of hits used in the reconstruction, and Number of track hits.}
		\label{events}
		\begin{tabular}{||c|c| c| c| c| c| c|c|c|c||} 
			\hline
			Cl	&	$E_{\rm{rec}}$ [TeV] & $\theta$ [$\degree$] & $\phi$ [$\degree$] & $\rho$ [m]&L &Q [p.e.]& nHits & nRecoHits & nTrackHits\\ [0.5ex] 
			\hline\hline
			
			0  & 79.7 & 71.30& 4.96 & 47.65   &1.05 &1665.01 &106 & 49 & 0\\
			\hline	
			4 & 46.6 & 43.11 &  247.31 &61.55 & 1.25 & 3804.82 &69 &23 &0\\
			\hline	
			
		\end{tabular}
	\end{center}
	\section{Conclusion}
	We have developed and optimized several suppression methods for background cascades produced along the muon tracks. Most of the  methods try to find  hits from the muon track in the event. These suppression methods are optimized with Monte Carlo (MC) simulations for season 2019. Subsequently, experimental data from season 2019 were processed and analyzed.
	A total of 25 contained and uncontained events fulfill the BDT cut condition to be  the potential candidates for neutrino cascade events. Among these events, two were reconstructed with high energies, 79.7 TeV for one contained event and 46.6 TeV for the uncontained.
	
	\section{Acknowledgements}
	
	The work was partially supported by RFBR grant 20-02-00400. The CTU group acknowledges the support by European Regional Development Fund-Project No. CZ.02.1.01/0.0/0.0/16\_019/0000766. The CU group acknowledges the support by the VEGA Grant Agency of the Slovak Republic under Contract No. 1/0607/20 and support by the  Grants  of  Comenius  University  in  Bratislava, Slovakia UK/231/2021. We also acknowledge the technical support of JINR staff for the computing facilities (JINR cloud).

	%
	%
	%
	

\begin{thebibliography}{99}
		\bibitem{GVD}
		https://baikalgvd.jinr.ru/
		
		\bibitem{clusterVolume}
		Baikal-GVD Collaboration: A.D. Avrorin et al., Search for cascade events with BaikalGVD, PoS-ICRC2019-0873, arXiv:1908.05430, doi:10.22323/1.358.0873
		
		\bibitem{IceCubeCascade}
		IceCube Collaboration: M. G. Aartsen et al., Atmospheric and astrophysical neutrinos above 1 TeV interacting in IceCube. Physical Review D 91, 022001 (2015), doi: 10.1103/PhysRevD.91.022001
		
		\bibitem{parallel}
		O. Tange, GNU Parallel - The Command-Line Power Tool, The USENIX Magazine (2011), doi: 10.5281/zenodo.16303
		
		
		\bibitem{ANTARES}
		ANTARES Collaboration: A. Albert et al., An algorithm for the reconstruction of neutrino-induced showers in the ANTARES neutrino telescope, 	arXiv:1708.03649 [astro-ph.IM], doi:10.3847/1538-3881/aa9709
		
		
		
		
	\end{thebibliography}
\end{document}